# The physics of magnetization

Yuri Mnyukh
*76 Peggy Lane, Farmington, CT, USA, e-mail: yuri@mnyukh.com*
(Dated: January 4, 2011)

Experimental observations recently reported in *Nature* are in accord with the concept that spin orientation is a permanent feature of the spin carrier. It follows that magnetization is realized by rearrangement of the crystal structure. The rearrangement occurs by the general *contact* nucleation-and-growth mechanism.

The experimental facts reported in *Nature* [1-3] are indicative of the need for changing the current interpretation of magnetization mechanism. Magnetization, whether it is caused by application of magnetic field or changing temperature, is presently believed to be a "rotation" ("switching", "reversal") of spins *in* the crystal which can remain intact. The new interpretation, put forward in 2001 [4], coherently accounts for all puzzling observations made in [1-3]. It states that *spin orientation is a permanent feature of the spin carrier* (atom, molecule), therefore magnetization inevitably involves turning the carriers. In other words, magnetization is inseparably linked with rearrangement of the crystal. Thus, in order to comprehend the real magnetization process, understanding of the molecular* mechanism of solid-state rearrangements is required.

The following is the general mechanism of structural rearrangements, deduced from the studies presented initially by the sequence of journal articles [5-18] and then summarized in the book [4].

• Rearrangements in a solid state are realized by crystal growth involving nucleation and propagation of interfaces. Neither ferromagnetic phase transitions (see below), nor ferroelectric phase transitions [19] are excluded from this rule. Not a single sufficiently well-documented example exists of this process being homogeneous (cooperative).

• The nuclei are located in specific crystal defects - microcavities of a certain optimum size. These defects contain information on the condition (*e.g.*, temperature) of their activation and the orientation of resultant crystal lattice. The nucleation can be epitaxial, in which case a certain orientation relationship between the initial and resultant structures is observed.

• The interface is a rational crystallographic plane of the resultant crystal lattice. It is named "contact interface" owing to direct molecular contact between the two lattices without any intermediate layer. The molecular rearrangement proceeds according to *edgewise* (or *stepwise*) mechanism (Fig.1) consisting of formation of "kinks" (steps) at the flat interface and filling them, molecule-by-molecule, until the layer is complete, and building successive layers in this manner.

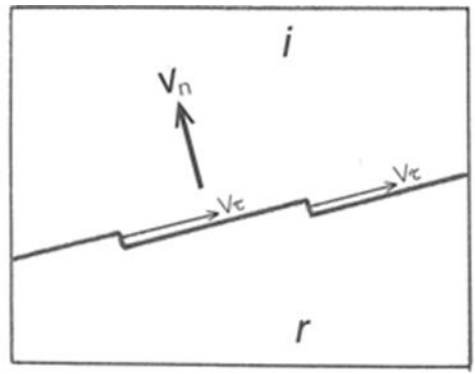

FIG. 1. The *edgewise* mechanism of phase transitions and any other rearrangements in solid state, such as at domain boundaries. The sketch illustrates the mode of advancement of interface in the **n** direction by shuttle-like strokes of small steps (kinks), filled by molecule-by-molecule, in the **τ** direction; *i* and *r* – are initial and resultant crystals, respectively.. (A crystal growth from liquids is realized by the same manner). More detailed description of the mechanism and its advantages is given in [4].

Lavrov *at al.* [1] (LKA) have observed crystal rearrangement of a antiferromagnet by magnetic field. Three relevant aspects of the LKA work will be highlighted:

*First.* In terms of the conventional science the phenomenon itself must not exist. As LKA indicated, "the common perception [is that] magnetic field affects the orientation of spins, but has little impact on the crystal structure." But the structure did changed in their experiments. According to the new concept, however, structural rearrangement is the only way of changing spin reorientation (i.e., of magnetization).

*Second.* LKA note that "one would least expect any structural change to be induced in antiferromagnet where spins are antiparallel and give no net moment". Nevertheless, such an unexpected phenomenon has took place. The reason for that seeming contradiction is rooted in the belief that spins in an antiferromagnet are strongly bound together by the Heisenberg's "exchange forces", therefore the external field H, which is weak in comparison, cannot deal separately with the parallel and antiparallel components of the spin system.

The legitimacy of the "exchange forces" theory was challenged in [4]. Even its initial verifications had to prevent its acceptance. The case in point is that the verifications have produced a *wrong sign* of the exchange forces. Despite this fatal defect, this theory was taken for granted. But Feynman [20] was skeptical at least, as seen from these statements: "When it was clear that quantum mechanics could supply a tremendous spin-oriented force - even if, apparently, of the wrong sign - it was suggested that ferromagnetism might have its origin in this same force", and "The most recent calculations of the energy between the two electron spins in iron still give the *wrong* sign", and "It is worse than that. Even if we *did* get the right sign, we would still have the question: why is a piece of lodestone in the ground magnetized?", and even "This physics of ours is a lot of fakery." The sign problem was carefully examined in a special review [21] and found fundamentally unavoidable in the Heisenberg model. It was suggested that the "neglect of the sign may hide important physics."

LKA actually dealt with the *antiferromagnet → ferromagnet* phase transition when every second *spin carrier* was turned, so that its spin turned toward the direction of external magnetic field. Evidently, spins were strongly bound to their carriers rather than to each other.

*Third*. LKA observed a generation and motion of crystallographic twin boundaries and kinks moving along them, resulting in a crystal rearrangement. While their findings are inconsistent with the idea of spin "switching" or with any cooperative phase transition, they are in accord with the illustrated in Fig.1 magnetization mechanism by crystal growth.

Novoselov *et al*. [2] recorded magnetization picture with a high resolution never before attained. They found that the ferromagnetic domain interface propagated by distinct jumps matching the lattice periodicity, the smallest being only a single lattice period. Some results also suggested that "kinks" were running along the interface. The authors interpreted the interface movements as following the Peierls potential of crystal lattice and stated that further theoretical and experimental work is needed to understand the unexpected dynamics of domain walls. The phenomenon, however, had been described, predicted to be traced to the molecular level, explained and illustrated with a molecular model in [8,11,13,17, 4] (see Fig.1 above). In fact, the same mode of interface propagation (running kinks and filling layer-by-layer) was observed by LKA [1] as well, only on a more macroscopic level, and the fact that this led to a real crystal rearrangement was firmly established.

Tudosa *et al*. [3] estimated experimentally the ultimate speed of "magnetization switching" in tiny single-domain particles - an important issue in developing of magnetic memory devices. The speed turned out three orders of magnitude lower than was predicted and, besides, was not the same in the effected particles. The error of that prediction is hidden in the term "switching", in other words, in the assumption of a cooperative spin rotation *in* the current crystal structure. The lower speed had to be expected if magnetization is not a "switching", but occurs by nucleation and growth in every individual domain. Nucleation is heterogeneous, requires specific crystal defects and not simultaneous in different particles. It is nucleation that controls re-magnetization of small single-domain particles [22].

The experimental observations presented in [1,2,3] provide strong evidence that magnetization is realized by structural rearrangement according to the specific *edgewise* mechanism involving nucleation and propagation of interfaces, rather than by spin rotation, switching or reversal in the same structure.

---------------------------------------------------

* "Atomic" and "molecular" are interchangeable in the text.